\documentclass[aps,prb,reprint,showpacs,groupedaddress,footinbib,citeautoscript]{revtex4-1}

\usepackage{amssymb}
\usepackage{amsmath}
\usepackage{amsfonts}
\usepackage{color}
\usepackage{graphicx}
\usepackage{bm}

\usepackage{array}
\newcolumntype {s}[1]{@{\hspace{#1}}} % space
\newcolumntype {R}{>{$}r<{$}}         % switch text <-> math
\newcolumntype {C}{>{$}c<{$}}         % switch text <-> math
\newcolumntype {L}{>{$}l<{$}}         % switch text <-> math

\usepackage{dcolumn}

\usepackage{wrapfig}

\graphicspath{{.}{./EPS/}}

\newcommand* {\vek}[1]{{\ensuremath{\bm{\mathrm{#1}}}}}
\newcommand* {\vekc}[1]{{\ensuremath{\bm{\mathcal{#1}}}}}
\newcommand* {\kk}{\vek{k}}
\newcommand* {\rr}{\vek{r}}

\newcommand* {\kcomp}{\kappa}
\newcommand* {\Ee}{\mathcal{E}}
\newcommand* {\kvek}{\bm{\kcomp}}

\begin{document}

\title{Magnetoelectric effect in bilayer graphene controlled by valley-isospin density}

\author{U. Z\"ulicke}
%\email{uli.zuelicke@vuw.ac.nz}
\affiliation{School of Chemical and Physical Sciences and MacDiarmid
Institute for Advanced Materials and Nanotechnology, Victoria
University of Wellington, PO Box 600, Wellington 6140, New Zealand}
\affiliation{Kavli Institute for Theoretical Physics, University of California, Santa Barbara,
CA 93106, USA}

\author{R. Winkler}
%\email{rwinkler@niu.edu}
\affiliation{Department of Physics, Northern Illinois University,
DeKalb, Illinois 60115, USA}
\affiliation{Materials Science Division, Argonne National
Laboratory, Argonne, Illinois 60439, USA}
\affiliation{Kavli Institute for Theoretical Physics, University of California, Santa Barbara,
CA 93106, USA}

\date{\today}

\begin{abstract}
  We show that bilayer graphene (BLG) exhibits magneto-electric (ME)
  effects that are formally similar to those commonly seen in band
  insulators with broken inversion and time-reversal
  symmetries. Three unusual features characterize the
  ME responses exhibited by BLG: (i)~unlike most other
  ME media, BLG is a conductor, (ii)~BLG has a
  non-quantized ME coupling even though its electronic structure
  \emph{does not\/} break parity and time-reversal symmetry, and
  (iii)~the magnitude of the ME coupling in BLG is
  determined by the valley-isospin density, which can be manipulated
  experimentally. This last property also enables a purely electric
  measurement of valley-isospin densities.
  While our theoretical arguments use BLG as an example, they
  are generally valid for any material with similar symmetries.
\end{abstract}


%\keywords{}

\maketitle

\section{Introduction and Overview of results}
 
Ordinarily when
matter is exposed to an electric field $\vek{\Ee}$ (a magnetic field
$\vek{B}$), an electric polarization $\vek{P}$ (magnetization
$\vek{M}$) is generated~\cite{jac99}. The magneto-electric (ME)
effect~\cite{ode70, fie05, ram07} refers to phenomena where
these common relations between applied field and resulting response
are crossed; i.e., when an applied electric field produces a finite
magnetization and a magnetic field results in an electric
polarization of the medium. The constitutive relations between the
microscopic fields $\vek{\Ee}$, $\vek{B}$ and the macroscopic
(induced) fields $\vek{D}$, $\vek{H}$ are then of the general form
\begin{subequations}\label{eq:constRel}
\begin{eqnarray}
  \vek{D} &\equiv& \epsilon_0\,\vek{\Ee} + \vek{P}
  = \epsilon_0\, \underline{\epsilon}
  \cdot\vek{\Ee} + \underline{\alpha} \cdot\vek{B} \quad , \\[1.0ex]
  \mu_0\, \vek{H} &\equiv& \vek{B} - \mu_0\, \vek{M}
  = \underline{\mu}^{-1} \cdot \vek{B} - \mu_0\, \underline{\alpha}
  \cdot\vek{\Ee} \quad .
\end{eqnarray}
\end{subequations}
Here $\underline{\epsilon}$, $\underline{\mu}$, and
$\underline{\alpha}$ are tensors describing the dielectric,
magnetic, and ME response, respectively. The ME tensor is usefully
written as~\cite{heh08, qi08, ess10}
\begin{equation}\label{eq:usualME}
\underline{\alpha} = \frac{\theta}{2\pi} \, \frac{e^2}{h} \, \openone +
\underline{\tilde \alpha} \quad ,
\end{equation}
where $\underline{\tilde\alpha}$ is traceless (and, in general, the
sum of a symmetric and an antisymmetric part), $e$ denotes the
elementary charge, and $h$ is the Planck constant.

Materials exhibiting ME coupling are currently attracting
significant interest~\cite{fie05, ram07, heh08, qi08,
ess10, li10a, oog12}, due to their unconventional basic
properties and also because the additional versatility that comes
from connecting electric and magnetic phenomena in new ways may
become the basis for useful device applications~\cite{ram07}.  Due
to the different symmetry properties of electric and magnetic
fields, ME coupling generally appears in materials in which
time-reversal and inversion symmetries are broken~\cite{fie05,
ram07}. Recently it was realized~\cite{qi08} that topological
insulators show a \emph{quantized\/} ME response ($\theta
\equiv \pi$, $\underline{\tilde\alpha}\equiv 0$). The analogies
between the electrodynamic phenomena in topological
insulators~\cite{qi08, moo10, qi11} and those arising in the context
of axion field theory~\cite{wil87} have opened up an intriguing link
between condensed-matter and elementary-particle
physics~\cite{fra08}.

Here we consider the electromagnetic properties of bilayer graphene
(BLG), which is an atomically thin conductor with unusual electronic
structure~\cite{mcc06, cas09, mcc13}. One of the interesting
features of charge carriers in BLG is that, in addition to their
ordinary (intrinsic) spin, electrons in BLG carry an orbital
pseudo-spin-1/2 degree of freedom that is rigidly locked with their
linear crystal momentum. Furthermore, states near the Fermi energy
exist in two valleys labelled $\vek{K}$ and $\vek{K'}$ that are
conveniently represented by a valley-isospin degree of freedom.
Several mechanisms have been proposed to separately address states
from the two valleys in BLG~\cite{yao08, xia07, mar08, abe09, sch10,
wu13, pra14} as part of the recent drive to establish a valleytronics
paradigm~\cite{ryc07, gun06, beh12}.

Our theoretical study reveals the existence of an unusual ME
coupling in BLG that is of the form
\begin{subequations}\label{eq:BLGmePar}
\begin{eqnarray}
\theta &=& 2\pi\, \frac{n_\mathrm{v}}{\bar n} \quad , \\
\underline{\tilde\alpha} &=& \frac{e^2}{h}\,
\frac{n_\mathrm{v}}{\bar n}\, \, \eta \left(
\begin{array}{rs{1.2em}rr} 1 & 0 & 0 \\ 0 & 1 &  0 \\ 0 & 0 & -2
\end{array} \right) \quad ,
\end{eqnarray}
\end{subequations}
where $n_\mathrm{v}$ is the valley-isospin density (i.e., the
difference of sheet densities for electrons from the two
valleys). Having a finite valley-isospin density corresponds to a
non-equilibrium many-particle state that has broken parity and
time-reversal symmetry, which enables the ME effect in a system
whose band structure has not broken these symmetries.~\cite{foot:nv}
Values for the BLG-specific parameters $\bar n$ and $\eta$
are given below. The dependence on $n_\mathrm{v}$ establishes an
intrinsic link between ME effects and valleytronics in BLG, creating
the possibility for electronic control of the generic ramifications
of ME coupling~\cite{wil87, qi08, fra08, kho13, fec14} that is not
available in other magnetoelectrics.  It is also remarkable in
the case of BLG that $\underline{\tilde\alpha}$ is finite and
$\theta$ non-quantized even though time reversal and parity
are proper symmetries of this material's bulk electronic structure.
As shown by a recent symmetry analysis~\cite{win12},
the possibility for unconventional couplings between electronic
degrees of freedom and electromagnetic fields in BLG arises due
to an unusual interplay of time-reversal and spatial symmetries.

\begin{figure}[tb]
\includegraphics[width=0.9\columnwidth]{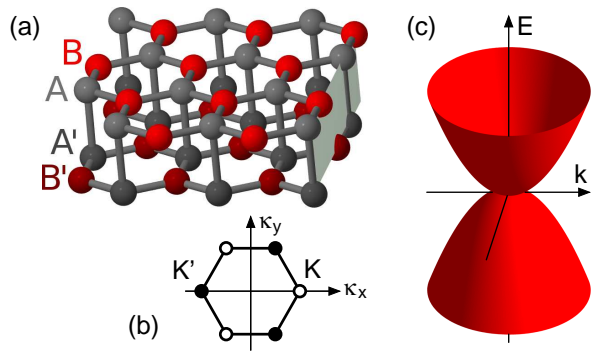}
\caption{\label{fig:blgStruct}
Basic structural and electronic properties of bilayer graphene. (a) Honeycomb
structure of a bilayer stack of graphene. Atoms in sublattice $A$ ($B$) are
marked in grey (red). A $yz$ plane is marked in light grey. (b) Brillouin zone
and its two inequivalent corner points $\vek{K}$ and $\vek{K}'$. The remaining
corners are related with $\vek{K}$ or $\vek{K}'$ by reciprocal lattice vectors.
(c) Dispersion $E(\kk)$ near the $\vek{K}$ point. We have $\kk \equiv \kvek -
\vek{K}$.}
\end{figure}

Using the well-understood electronic structure of BLG as a specific example,
our work establishes a more general paradigm of novel ME phenomena.
Indeed, our symmetry-based approach is valid beyond BLG for any materials
with similar symmetries. It may guide future research to identify, or even design,
materials exhibiting these symmetries that will then also display the ME effects
discussed here.

\section{Magneto-electric coupling in BLG}

BLG consists of two
layers of graphene stacked as shown in
Fig.~\ref{fig:blgStruct}(a). Near the $\vek{K}$ point in the
Brillouin zone [cf.\ Fig.~\ref{fig:blgStruct}(b)], the low-energy band
structure is described by an effective Hamiltonian~\cite{mcc06,
cas09, mcc13}
\begin{equation}\label{eq:basicBLGH}
  \begin{array}[b]{r>{\displaystyle}l}
\mathcal{H}^{\vek{K}} (\kk) = &
\frac{\hbar^2}{2 m_0}\left[ -u
\left(k_+^2 \, \sigma_+ + k_-^2 \, \sigma_-\right) + w\, k^2
\sigma_0  \right] \\[1.5ex] & \hspace{6em}
- \, \hbar v \left( k_- \, \sigma_+ + k_+ \, \sigma_-\right ) \; ,
  \end{array}
\end{equation}
where $\hbar\equiv h/(2\pi)$, $\vek{k}\equiv(k_x, k_y)$ is the
electrons' wave vector measured from $\vek{K}$, the Pauli matrices
$\sigma_{x,y,z}$ are associated with the sublattice (or,
equivalently, the layer-index) pseudospin degree of freedom,
$\sigma_0$ is the $2\times 2$ unit matrix, $\sigma_\pm= (\sigma_x
\pm i\sigma_y)/2$, and $k_\pm=k_x\pm i k_y$. Numerical values for
the effective-mass parameters $u$, $w$ and the speed $v$ are well
known~\cite{mcc06, cas09, mcc13}. Very close to the $\vek{K}$ point,
the energy dispersion resulting from Eq.\ (\ref{eq:basicBLGH})
mimics that of massless Dirac electrons, as is the case in
single-layer graphene.  However, as $u \gg w$, the dominant behavior
of electrons in BLG is captured by the quadratic dispersion shown in
Fig.~\ref{fig:blgStruct}(c).  To obtain the corresponding effective
Hamiltonian for electrons in the $\vek{K'}=-\vek{K}$ valley, we use
the relation~\cite{win10a} $\mathcal{H}^{\vek{K'}} (\kk) =
\mathcal{H}^{\vek{K}} (R_y^{-1}\kk)$, where $R_y$ denotes a mirror
reflection at the $yz$ plane [see Fig.~\ref{fig:blgStruct}(a)].

The band structure of BLG turns out to be strongly affected by
electric and magnetic fields~\cite{mcc06, mcc06b, oht06, min07,
cas07, zha09, nak09, kos10a, zha11, win12}.  Combining
the electronic degrees of freedom from the two valleys and using a
straightforward notation in terms of valley-isospin Pauli matrices
$\tau_0$ and $\tau_z$, the effective Hamiltonian becomes
$\mathcal{H} = \mathcal{H}_\mathrm{orb} + \mathcal{H}_\mathrm{pss} +
\mathcal{H}_\mathrm{ME}$, with
\begin{subequations}
\begin{eqnarray}
\mathcal{H}_\mathrm{orb} &=& \sum_{\tau=\pm 1} \frac{\tau_0 +\tau\, \tau_z}{2}
\,\, \mathcal{H}^{\tau\vek{K}}(\kk+e\vek{A}) - e\, \Phi \, \tau_0 \,\, , \\
\mathcal{H}_\mathrm{pss} &=& \left(\frac{g_\mathrm{e}}{2}\, \frac{\Ee_z}{c}\,
\tau_0 - \frac{g_\mathrm{m}}{2}\, B_z\, \tau_z \right) \mu_\mathrm{B}\,
\sigma_z \quad , \\ \label{eq:me_scalar}
\mathcal{H}_\mathrm{ME} &=& \left( \xi_\|\, \vek{\Ee}_\|
\cdot \vek{B}_\| + \xi_z\, \Ee_z B_z \right) e\, \tau_z \quad .
\end{eqnarray}
\end{subequations}
Here $\vek{\Ee}_\| = (\Ee_x, \Ee_y)$ and $\vek{B}_\| = (B_x, B_y)$,
and $c$ is the speed of light in vacuum. The part
$\mathcal{H}_\mathrm{orb}$ contains the coupling of planar electron
motion in the BLG sheet to the electromagnetic scalar and vector
potentials $\Phi(\rr)$ and $\vek{A}(\rr)$, respectively, which
satisfy $\vek{\nabla}\Phi + \partial_t \vek{A} = -\vek{\Ee}_\|$ and
$\vek{\nabla}\times\vek{A} = B_z\,\vek{\hat z}$. The effect of external
fields on the pseudo-spin degree of freedom is captured, in lowest
order, by $\mathcal{H}_\mathrm{pss}$, which accounts for the
pseudo-spin Zeeman splitting~\cite{zha11} and electric-field-induced
bandgap~\cite{mcc06, mcc06b, oht06, min07, cas07, zha09}. In the
following, we focus on the ramifications of
$\mathcal{H}_\mathrm{ME}$.

The term $\propto \Ee_z B_z$ has been discussed
previously~\cite{nak09, kos10a, zha11} in the context of
valley-contrasting magnetic moments. The existence of the
complementary coupling of in-plane electric and magnetic field
components $\propto \vek{\Ee}_\| \cdot\vek{B}_\|$ was recently
established by an invariant expansion for the BLG band
structure~\cite{bir74, win12}, which also revealed the origin of
such unconventional terms.  Firstly, as in single-layer
graphene~\cite{man07, win10a}, the two valleys are linked by
time-reversal and spatial symmetries in a \emph{combined} way
\cite{bir74}, which results in an unusual constraint on the
Hamiltonian describing the intravalley dynamics.  From a symmetry
perspective, this is the origin for the emergence of
massless-Dirac-fermion-like charge carriers in graphitic
materials~\cite{man07, win10a}. This constraint is likewise
satisfied by the terms appearing in
$\mathcal{H}_\mathrm{ME}$. Secondly, and unlike single-layer
graphene, the $\vek{K}$ point in BLG is characterized by a point
group ($D_3$) that does not distinguish between polar and axial
vectors as it only contains rotations as symmetry elements. As a
result, components of the electric and magnetic fields are not
distinguishable by symmetry and can therefore couple in normally
forbidden combinations to electronic degrees of freedom and to each
other~\cite{win12}.

\section{Lagrangian for electromagnetic fields in BLG}

The terms
appearing in $\mathcal{H}_\mathrm{ME}$ of Eq.~(\ref{eq:me_scalar})
are reminiscent of the $\vek{\Ee}\cdot\vek{B}$ contribution to the
axion Lagrangian~\cite{wil87, qi08}. They also have a valley
dependence. Assuming independent dynamics for electrons from the two
valleys, we can define the total sheet density $n_\mathrm{s} =
n^{\vek{K}}+n^{\vek{K}'}$ and particle-current density
$\vek{j}_\mathrm{s} = \vek{j}^{\vek{K}} + \vek{j}^{\vek{K}'}$ for
electrons from the two valleys as well as the differences
$n_\mathrm{v} =n^{\vek{K}} - n^{\vek{K}'}$ and $\vek{j}_\mathrm{v} =
\vek{j}^{\vek{K}} - \vek{j}^{\vek{K}'}$ as relevant
variables.~\footnote{The sheet densities and current densities are
obtained by integrating the corresponding three-dimensional
electronic densities over the $z$ coordinate perpendicular to the
BLG plane. Also, in this work, components of the electric and
magnetic field represent averages of these quantities over the
finite thickness of the BLG sample.}  They satisfy the continuity
equations
\begin{equation}
\partial_t\, n_\mathrm{s/v} + \vek{\nabla}\cdot \vek{j}_\mathrm{s/v} = 0
\quad .
\end{equation}

Application of the usual formalism~\footnote{See, e.g., E. Fradkin,
\textit{Field Theories of Condensed Matter Physics\/}, 2nd
ed. (Cambridge University Press, Cambridge, UK, 2013).  A related
calculation for BLG subject to crossed electric and magnetic fields
was recently performed by M. Katsnelson, G. Volovik, and M. Zubkov,
Ann. Phys. (NY) \textbf{331}, 160 (2013).} yields the
electromagnetic part of the BLG Lagrangian density as
\begin{subequations}
\begin{equation}
  \mathcal{L}_\mathrm{elm} = \mathcal{L}_\mathrm{Maxwell}
  + \mathcal{L}_\mathrm{min} + \mathcal{L}_\mathrm{ME} \quad .
\end{equation}
Here
\begin{equation}
  \mathcal{L}_\mathrm{Maxwell}
  =  \frac{\epsilon_0}{2} \, \vek{\Ee}\cdot\underline\epsilon\cdot\vek{\Ee}
      - \frac{1}{2\mu_0}\, \vek{B}\cdot \underline{\mu}^{-1} \cdot\vek{B}
\end{equation}
is the familiar Lagrangian for 3D Maxwell electrodynamics in an
ordinary medium with, in general, spatially varying dielectric and
magnetic tensors $\underline\epsilon$ and $\underline\mu$, and
\begin{equation}
  \mathcal{L}_\mathrm{min} = - e\, \big(\vek{j}_\mathrm{s}\cdot \vek{A} -
  n_\mathrm{s} \, \Phi\big)\, \delta(z)
\end{equation}
is the minimal coupling of the electromagnetic vector and scalar
potentials to the electric charge and current densities for
electrons in BLG. The part
\begin{equation}\label{eq:axionLag}
\mathcal{L}_\mathrm{ME} = - e\, n_\mathrm{v} \left( \xi_\parallel \,
\vek{\Ee}_\|\cdot \vek{B}_\| + \xi_z \, \Ee_z \, B_z \right) \delta(z)
\end{equation}
\end{subequations}
contains the axion-like ME coupling arising from the unconventional
contribution (\ref{eq:me_scalar}) to the BLG Hamiltonian.  The fact
that the valley-isospin density $n_\mathrm{v}$ couples
to the field combinations in Eq.~(\ref{eq:axionLag}) is reasonable
as $n_\mathrm{v}$ is odd under both time-reversal and parity
transformations. The terms shown in Eq.~(\ref{eq:axionLag}) are
therefore allowed by the symmetry operations applicable to the
Lagrangian for the electromagnetic field.

The existence of $\mathcal{L}_\mathrm{ME}$ leads to modifications of
the inhomogeneous Maxwell's equations~\cite{wil87} (i.e., Gauss's
and Amp\`ere's laws), resulting in constitutive relations of the
form given in Eq.~(\ref{eq:constRel}) with
Eq.~(\ref{eq:usualME}). Straightforward calculation yields $\theta$
and $\tilde\alpha$ as given in Eqs.~(\ref{eq:BLGmePar}), with
\begin{equation}\label{eq:MEnPar}
\bar n = \frac{e}{h}\, \frac{3\, d_\mathrm{eff}}{\xi_z + 2 \xi_\|}
\quad, \hspace{2em}
\eta = \frac{\xi_\| - \xi_z}{\xi_z + 2 \xi_\|} \quad .
\end{equation}
Here $d_\mathrm{eff}$ denotes the effective electronic width of BLG
in the perpendicular ($z$) direction.

\section{Magneto-electric response in BLG}

The modified
constitutive relations (\ref{eq:constRel}) imply the existence of
extra contributions to the total sheet density $n_\mathrm{s}$ and
current $\vek{j}_\mathrm{s}$.~\footnote{In the previously encountered
situations where the ME medium is an insulator, these are associated
with bound charges. In BLG, they correspond to free charges.} In
terms of the parameters governing ME responses in BLG, these can
be written as
\begin{subequations}
\label{eq:nj_intermed}
\begin{eqnarray}
  n_\mathrm{s,ME}  &=&  - \vek{\nabla}\cdot\left[ n_\mathrm{v} \left(
  \xi_\parallel\, \vek{B}_\| + \xi_z \, B_z\,\vek{\hat z}\right)\right] , \\
  \vek{j}_\mathrm{s,ME} &=& \vek{\nabla} \times\left[ n_\mathrm{v}
    \left( \xi_\parallel\, \vek{\Ee}_\| + \xi_z \, \Ee_z\vek{\hat z}\right)
  \right]
  \nonumber \\ \label{eq:j_intermed} & & \hspace{4em} {}
  +\partial_t \left[ n_\mathrm{v} \left( \xi_\parallel\,
  \vek{B}_\| + \xi_z \, B_z\,\vek{\hat z} \right) \right] .
  \hspace{2em}
\end{eqnarray}
\end{subequations}
Obviously, these are nonzero only in case of spatial or temporal
variations of the quantities appearing in square brackets in
Eqs.~(\ref{eq:nj_intermed}). For example, the BLG material
parameters $\xi_\|$ and $\xi_z$ are nonzero within BLG but they
vanish outside the sample. Thus both their change at the sample
boundary as well as any spatial variations of $n_\mathrm{v}$,
$\vekc{E}$ or $\vek{B}$ result in an induced charge density and/or
current. We then have
\begin{equation}
n_\mathrm{s,ME} =  - \left[ \vek{\nabla}_\| \left( n_\mathrm{v}\,
\xi_\parallel \right)\right]\cdot\vek{B}_\|
- n_\mathrm{v}\, \xi_\parallel\,
\vek{\nabla}\cdot\vek{B}_\| - n_\mathrm{v}\, \xi_z \,\partial_z B_z  .
\end{equation}
Using the homogeneous Maxwell equation $\vek{\nabla}\cdot\vek{B}=0$,
the magnetic-field-induced charge density is found to be
\begin{equation}\label{eq:axion_rho}
n_\mathrm{s,ME}  =  - \vek{B}_\| \cdot \vek{\nabla} \left( n_\mathrm{v}\,
\xi_\parallel \right) - n_\mathrm{v} \left( \xi_\parallel - \xi_z \right)
\vek{\nabla}_\| \cdot \vek{B}_\| \quad .
\end{equation}
Similarly, inserting the homogeneous Maxwell equation $\vek{\nabla}
\times\vek{\Ee}= -\partial_t\vek{B}$ into Eq.~(\ref{eq:j_intermed})
yields~\footnote{In arriving at Eq.~(\ref{eq:axion_j}), the term
\begin{displaymath}
  \hspace{2em}
  \vek{\Ee}_\parallel\times \vek{\nabla}
  \left( n_\mathrm{v}\, \xi_\parallel \right)
  - \left[B_z\, \partial_t \left( n_\mathrm{v}\, \xi_z \right)
    + n_\mathrm{v} \left( \xi_z - \xi_\parallel \right)
    \partial_t B_z \right] \vek{\hat z}
\end{displaymath}
has been omitted for two reasons.  Firstly, $\vek{j}_\mathrm{s}$ can
only have in-plane components. Secondly, and more importantly, the
above contribution arises due to an \emph{in-plane electric\/} or
\emph{perpendicular magnetic\/} field, both of which strongly affect
the orbital motion of electrons in the BLG sheet. In contrast, the
presence of \emph{perpendicular electric\/} or \emph{in-plane
magnetic\/} fields will not lead to drastic orbital effects and,
therefore, reveal clearly the ME effects discussed here.}
\begin{eqnarray}\label{eq:axion_j}
\vek{j}_\mathrm{s,ME} &=& \left[ \Ee_z \vek{\nabla} \left( n_\mathrm{v}
\, \xi_z\right) + n_\mathrm{v}\left( \xi_z - \xi_\parallel \right) \vek{\nabla}
\Ee_z \right] \times \vek{\hat z} \nonumber \\ && \hspace{3.5cm}
+ \, \vek{B}_\parallel \, \partial_t
\left( n_\mathrm{v} \, \xi_\parallel \right).
\quad
\end{eqnarray}

\begin{figure}[t]
  \includegraphics[width=1.0\columnwidth]{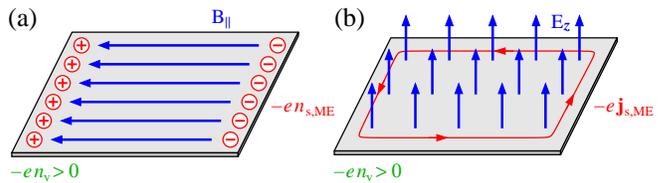}
  \caption{\label{fig:axion}
  Illustration of magneto-electric effects in BLG. If the valley-isospin
  density $n_\mathrm{v}$ is finite, application of an in-plane magnetic
  (perpendicular electric) field as shown in panel (a) [(b)] induces
  charges (currents) at the sample boundary.}
\end{figure}

Equations~(\ref{eq:constRel}), (\ref{eq:axion_rho}) and
(\ref{eq:axion_j}) summarize the ME effects in BLG. Charge
\emph{densities\/} are induced by in-plane magnetic fields where
$n_\mathrm{v}$ varies spatially, especially at the BLG sheet's
boundaries in the $xy$ plane; and a perpendicular electric field
generates in-plane \emph{currents\/} flowing perpendicular to the
spatial gradient of $n_\mathrm{v}$, especially at the system's
boundaries in the $xy$ plane.  See Fig.~\ref{fig:axion} for an
illustration. Furthermore, in the likely case~\footnote{The value of
$\xi_{\|}$ is currently not known.} that $\xi_\|\ne \xi_z$, currents
(charge densities) are induced in the \emph{bulk\/} of BLG by
spatial inhomogeneity of the perpendicular electric (in-plane
magnetic) field component(s) wherever $n_\mathrm{v}$ is finite, and
bulk currents are generated by an \emph{in-plane\/} magnetic field
when $n_\mathrm{v}\, \xi_\parallel$ is time-dependent.

Formally, the ME response of BLG turns out to be analogous to that
of a uniaxial ME medium such as Cr$_2$O$_3$~\cite{heh08, fec14}.  In
particular, the features arising from spatial inhomogeneity of the
quantities $n_\mathrm{v} \,\xi_\parallel$ and $n_\mathrm{v}\, \xi_z$
are similar to those associated with axion electrodynamics in
3D~\cite{wil87, qi08}. Given that BLG and ordinary magnetoelectrics
belong to very different materials classes (Cr$_2$O$_3$ is an
antiferromagnetic insulator), this phenomenological similarity is
quite surprising.  The crucial difference between the two cases is
illustrated by the ME parameters.  Whereas $\theta$ and
$\underline{\tilde\alpha}$ are fixed in typical ME materials, their
values depend on the valley-isospin density $n_\mathrm{v}$ in
BLG.

Generation of a finite $n_\mathrm{v}$ is possible by various means,
including valley-filter contacts~\cite{ryc07, xia07, mar08, abe09,
sch10, wu13, pra14}, optical excitation~\cite{yao08, mak14}, or
using the ME coupling as a source for valley
polarization~\cite{xia07, kos10a, zha11} (cf.\ discussion below). In
the experimentally accessible situation where inter-valley
scattering in the relevant part of the sample is weak, the resulting
non-equilibrium state with $n_\mathrm{v} \ne 0$ will be long-lived
and enables the observation of ME effects even nonlocally (i.e.,
away from the region where $n_\mathrm{v}$ is created).

\section{Magnitude and application of magneto-electric responses}

The Slonczewski-Weiss-McClure (SWM) model~\cite{mcc57} applied to
BLG yields in terms of the SWM tight-binding parameters
(see also Refs.~\onlinecite{kos10a, zha11})
\begin{equation}\label{eq:MEprefac}
  \xi_z = \frac{\gamma_0^2}{\gamma_1^2} \,
  \frac{3\, g_\mathrm{e}}{8} \, \frac{e}{\hbar} \,
  \frac{\hbar a^2}{m_0 c}
  \quad .
\end{equation}
Recent experiments~\cite{zha09} demonstrated that a vertical
electric field of $\sim 1$~V/nm generates a bandgap of $\sim 0.1$~eV
in BLG, implying $g_\mathrm{e}\approx 500$ consistent with
first-principles calculations~\cite{kon12}. Using the values
$\gamma_0 = 3.0$~eV, $\gamma_1 = 0.32$~eV,
and $a = 0.245$~nm, we find $\xi_z \approx 6 \times
10^{-4}~\mathrm{nm / T}$. Assuming furthermore $\xi_\|\approx \xi_z$
and $d_\mathrm{eff} = 0.1$~nm~\cite{kon12}, the parameter
$\bar n \approx 4 \times 10^{12}~\mathrm{cm}^{-2}$ is obtained from
Eq.~(\ref{eq:MEnPar}). Thus valley-isospin densities
$n_\mathrm{v} \gtrsim 10^{10}\, \mathrm{cm}^{-2}$ are expected to
generate observable ME effects.

To estimate the magnitude of boundary charges and currents arising
from ME coupling in BLG, we use Eqs.~(\ref{eq:axion_rho}) and
(\ref{eq:axion_j}) to find
\begin{subequations}
\begin{eqnarray}
n_\mathrm{s,ME}  &=& - B_\|^{(\perp)}\, \frac{d (n_\mathrm{v}\, \xi_\|
)}{d r_\perp} \quad , \\
j_\mathrm{s,ME} &=& \Ee_z \, \frac{d (n_\mathrm{v} \, \xi_z)}{d r_\perp}
\quad , \\
I_\mathrm{ME} &\equiv& -e \int dr_\perp \,\,\, j_\mathrm{s,ME}
= -e\, \Ee_z \, \xi_z\,  n_\mathrm{v} \quad . \quad
\end{eqnarray}
\end{subequations}
Here $r_\perp$ is the in-plane coordinate perpendicular to a chosen
boundary of the BLG sample, $B_\|^{(\perp)}$ denotes the in-plane
magnetic-field component in that perpendicular direction, and
$I_\mathrm{ME}$ is the total charge current flowing parallel to the
boundary. Denoting by $l$ the length scale over which the quantity
$n_\mathrm{v}\,\xi_\|$ changes at the BLG-sample boundary and
assuming $\xi_\| \approx \xi_z\sim 6\times 10^{-4} \, \mathrm{nm/T}$,
we find
\begin{subequations}
\begin{eqnarray}\label{eq:MEcharge}
n_\mathrm{s,ME} &\sim&  B_\|^{(\perp)} \, \frac{\xi_\|}{l} \,
n_\mathrm{v} \sim 6 \times 10^{-4} \,\, \frac{B_\|^{(\perp)}
[\mathrm{T}]}{l [\mathrm{nm}]} \,\, n_\mathrm{v} \, , \quad \\
\label{eq:MEcurrent} I_\mathrm{ME} &\sim& 10\, \mathrm{nA} \,\,\,
\Ee_z \!\left[\frac{\mathrm{V}}{\mathrm{nm}}\right] \,\, n_\mathrm{v}\left[
10^{10}\,\mathrm{cm}^{-2}\right]  \, .
\end{eqnarray}
\end{subequations}
According to Eq.~(\ref{eq:MEcharge}), magnetic-field-induced
boundary charges are very small for realistic field magnitudes and
valley-isospin densities. In contrast, the boundary currents
that can be created via the ME effect are definitely measurable,
according to Eq.~(\ref{eq:MEcurrent}). Detection of these boundary
currents can thus be used as a direct measurement of a finite
valley-isospin density.

\section{Magneto-electric coupling as a source of valley-isospin
density}

As shown above, charges (currents) are induced in
the presence of a finite valley-isospin density and a magnetic
(electric) field by virtue of the ME coupling. In short, a finite
$n_\mathrm{v}$ plus \emph{one type\/} of field give rise to a
total-charge or current response. These effects embody the emergent
electromagnetism in BLG.  Conversely, the ME coupling affects the
electronic structure of a BLG sample in the \emph{simultaneous
presence of parallel electric and magnetic fields\/}, introducing a
uniform energy shift for electron levels that is opposite in the two
valleys. It is clear that such an effect must introduce a
valley-isospin density, as electrons redistribute from one valley
to the other in order to keep a constant Fermi energy throughout the
system. Initially, only the term $\propto \Ee_z\, B_z$ was known, and
a careful treatment of it (taking account of the orbital effects of $B_z$,
i.e., Landau quantization) indeed found that a valley-isospin density
is generated~\cite{xia07, kos10a, zha11}.

The term $\propto \xi_\| \, \vek{\Ee}_\|\cdot \vek{B}_\|$
imposes the same type of energy shift and thus also induces a
valley-isospin density, which we can estimate to
be~\footnote{The valley-asymmetric energy shifts $\propto \Ee_z\, B_z$,
\protect{$\vek{\Ee}_\| \cdot \vek{B}_\|$} are part of the effective
low-energy-Hamiltonian description that applies only
at energies well below the interlayer coupling energy
$\gamma_1\approx 0.32 \,$eV.  Hence there exists an upper bound
$n_\mathrm{v} \ll 2 m_0 \gamma_1/(\pi\hbar^2 u)\approx 8\times
10^{12}\,$cm$^{-2}$ to the valley-isospin density that can be
generated by this mechanism.}
\begin{equation}\label{eq:estNV}
n_\mathrm{v} = \frac{2}{\pi}\, \frac{e\, m_0}{\hbar^2} \,
\frac{\xi_\|}{u} \, \Ee_\|\, B_\|
\sim \Ee_\| \! \left[ \frac{V}{\mathrm{mm}} \right]\, B_\| [T] \times
10^{12}\,\mathrm{cm}^{-2} .
\end{equation}
Here we assumed $\xi_\| \sim \xi_z \approx 6 \times
10^{-4}~\mathrm{nm / T}$ to obtain the numerical value given on
the r.h.s.\ of Eq.~(\ref{eq:estNV}).
In contrast to the situation involving field components
perpendicular to the sheet, the magnetic field $\vek{B}_\|$ has no
orbital consequences.  However, $\vek{\Ee}_\|$ couples
electrostatically to the \emph{total\/} charge density and gets
screened from the interior of a conducting BLG sample. Thus, it will
at most induce a finite $n_\mathrm{v}$ at the boundary over a
distance of the screening length.

\section{Conclusions}

Fundamental symmetry considerations reveal
the presence of a valley-contrasting magneto-electric coupling
(\ref{eq:me_scalar}) for electrons in bilayer graphene~\cite{xia07,
kos10a, zha11, win12}.  As a result, the Lagrangian
governing the co-operative dynamics of electrons and electromagnetic
fields has a contribution where the valley-isospin density
couples to the parity- and time-reversal-odd terms $\propto
\vek{\Ee}_\| \cdot \vek{B}_\|$ and $\Ee_z B_z$ [cf.\
Eq.~(\ref{eq:axionLag})]. The resulting non-quantized
magneto-electric response of bilayer graphene mirrors that of
conventional uniaxial magneto-electric media, which is unusual
because the latter are typically insulating materials whose
electronic degrees of freedom exhibit broken parity and time-reversal
symmetry. \cite{foot:nv}
The unexpected behavior of electrons in graphene bilayers arises
because constraints due to invariance under time reversal and parity
involve the two valleys, rendering the dynamics of electrons within
individual valleys qualitatively different from that of ordinary charge
carriers. In particular, the intra-valley dynamics in bilayer graphene
cannot distinguish between polar and axial vectors \cite{win12} and,
as a result, supports the magneto-electric coupling discussed here.
As the validity of our symmetry-based approach is not restricted to
bilayer graphene, other materials with similar properties are likely to
exist or become available in the future.

Linking valleytronics with magneto-electric effects establishes new
pathways for each of these fields.  For example, the dependence of
the magneto-electric effects in bilayer graphene on the valley-isospin density
allows one to detect such charge imbalances between the valleys, e.g.,
via the boundary currents generated by a perpendicular electric field.
Also, the unusual electrodynamic properties exhibited by
magnetoelectrics~\cite{kho13, fec14} become tunable in bilayer graphene by
controlling its valley-isospin densities and currents.

\begin{acknowledgments}
This research was supported by the NSF
under Grant No.\ DMR-1310199 and, at KITP, by Grant No.\
PHY11-25915.  Work at Argonne was supported by DOE BES under
Contract No. DE-AC02-06CH11357.  Useful discussions with
J.~J. Heremans, A.~H.  MacDonald, and I.~Martin are gratefully
acknowledged.
\end{acknowledgments}

%\bibliography{axion}
%merlin.mbs apsrev4-1.bst 2010-07-25 4.21a (PWD, AO, DPC) hacked
%Control: key (0)
%Control: author (8) initials jnrlst
%Control: editor formatted (1) identically to author
%Control: production of article title (-1) disabled
%Control: page (0) single
%Control: year (1) truncated
%Control: production of eprint (0) enabled
%

\end{document}